\documentclass[aps,prc,npa,preprint,groupedaddress]{revtex4}

\usepackage{youngtab}

\usepackage{epsfig}

\begin{document}

\title{\boldmath The $qqqq\bar{q}$ components and the magnetic moments
of the charmed and the bottomed baryons}

\author{C. S. An}
\email[]{ancs@mail.ihep.ac.cn} \affiliation{Institute of High Energy
Physics, CAS, P. O. Box 918, Beijing 100049, China,}
\affiliation{Graduate University, Chinese Academy of Sciences,
Beijing 100049, China}

\thispagestyle{empty}

\date{\today}

\begin{abstract}

We give the explicit wave functions of the $qqqq\bar{q}$ components
of the $C=+1$, $J=1/2$ charmed baryons, $\Sigma_{c}$, $\Lambda_{c}$
and $\Xi_{c}^{a}$, and calculate the magnetic moments by adding the
5q components contributions, and we also compute the magnetic
moments of the $\Sigma_{b}$ and $\Sigma_{b}^{*}$ baryons. The
influence of the additional light and strange $q\bar{q}$ pairs is
investigated. As we know, the constituent quark masses of the charm
and beauty quarks are much heavier than that of the light and
strange quarks, consequently, the hidden flavor contributions to the
baryons magnetic moments may be significant. What is interesting is
that the inclusion of the $qqqq\bar{q}$ components contributions
leads to different $\Lambda^{+}_{c}$, $\Xi^{a+}_{c}$ and
$\Xi^{a0}_{c}$ magnetic moments, all of which are predicted to be
the same value $0.38\mu_{N}$ on the basis of the classical $qqq$
quark model. And it's shown that the differences of these magnetic
moments are independent of the constituent mass of the charm quark.

\end{abstract}

\maketitle

\section{introduction}
\label{sec:1}

Recently, the measurements of the $\bar{d}/\bar{u}$ asymmetry in the
nucleon sea indicate a considerable isospin symmetry breaking in the
light quark sea of the nucleon. This indicates that the nucleon
contains notable $qqqq\bar{q}$ components, if no more exotic
components, besides the conventional $qqq$
component\cite{towell,NA51,NMC,HERMES}. On the other hand, some
experiments on parity violation in electron-proton scattering
moreover indicate that the $s\bar{s}$ quark pairs lead to nonzero
contributions to the magnetic moment of the
proton\cite{happex1,happex2,sample1,sample2,a41,a42,g0}. The
experimental results on the strangeness magnetic moments can be
described , at least qualitatively, by $uuds\bar{s}$ configurations
in the proton with the anti-quark is in its S-state and the four
quarks subsystem in its P-state\cite{zou1,zou2,zou3}. In Ref.
\cite{liu}, taking large admixture of the pentaquark components to
the conventional $qqq$ components, a natural explanation for the
mass ordering of the $N^{*}(1440)1/2^{+}$, $N^{*}(1535)1/2^{-}$ and
$\Lambda^{*}(1405)1/2^{-}$ resonances has been given, which is very
puzzling in the classic $qqq$ quark model.

In Ref.\cite{bomm}, the hidden flavor contributions to the baryon
octet and decuplet magnetic moments have been investigated, and the
results show that the average deviation from the empirical magnetic
moment values drops from $\sim9\%$ of the static quark model
predictions to $\sim5\%$ when the contributions of the $qqqq\bar{q}$
components are taken into account. The improvement is most notable
in the case of the $\Sigma^{-}$($\sim14\%$ to $\sim4\%$),
$\Xi^{0}$($\sim21\%$ to $\sim3\%$) and $\Lambda$($\sim10\%$ to
$\sim0\%$). All these fact above suggest that there are significant
$qqqq\bar{q}$ components in baryons.

In the case of the charmed and bottomed baryons, the masses are much
heavier than that of the $SU(3)$ octet and decuplet, such as that
the mass of the $\Lambda_{c}$($J^{(P)}=\frac{1}{2}^{(+)}$) baryon is
about 2286 MeV, the $\Sigma_{c}$($J^{(P)}=\frac{1}{2}^{(+)}$) is
about 2455 MeV \cite{pdg}, and the new measured bottomed baryons
$m_{\Sigma_{b}^{-}}=5816$ MeV, $m_{\Sigma_{b}^{+}}=5808$ MeV,
$m_{\Sigma_{b}^{*-}}=5837$ MeV and $m_{\Sigma_{b}^{*+}}=5829$ MeV
\cite{fermilab}, which may result in that the $qqqq\bar{q}$
components play a more important role in the charmed and bottomed
baryons. Because of that the charm and beauty quarks are much
heavier than the light and strange quarks, the $qqqq\bar{q}$
components, i.e. the hidden light and strange flavors may have much
more significant contributions to the baryon magnetic moments. Here
the contributions to the $\Lambda_{c}^{+}$, $\Sigma_{c}$ and
$\Xi_{c}^{a}$ as well as the $\Sigma_{b}$ and $\Sigma_{b}^{*}$
baryons magnetic moments from the $qqqq\bar{q}$ wave function
components with at most one $u\bar{u}$, $d\bar{d}$ or $s\bar{s}$ sea
quark pair are calculated in the non-relativistic constituent quark
model. For the sake of the much heavier masses of the charm and
beauty quarks, we neglect the contributions of the $c\bar{c}$ and
$b\bar{b}$ components.

What is interesting is that because the light and strange degrees of
freedom are in spin zero configuration for the $\Lambda_{c}$,
$\Xi_{c}^{a+}$ and $\Xi_{c}^{a0}$ in the classic $qqq$ quark model,
all the magnetic moments of these baryons come from the
contributions of the charm quark, if the constituent mass of the
charm quark is set to be $m_{c}=1.7$ GeV, these magnetic moments
will be $0.38\mu_{N}$. But the magnetic moments of these baryons
should be different when the contributions of the $qqqq\bar{q}$
components in these baryons are taken into account, for the
different flavor-spin configurations of the 5q components in these
baryons. In this manuscript, it's shown that the differences are
independent of the constituent mass of the charm quark.

The present manuscript is organized in the following way. In Section
\ref{sec:2} the explicit flavor wave functions of the $qqqq\bar{q}$
components in the charmed and bottomed baryons are given. Section
\ref{sec:3} contains the expressions and the corresponding numerical
results for the charmed baryons. The corresponding expressions and
numerical results for the $\Sigma_{b}$ and $\Sigma_{b}^{*}$ magnetic
moments are given in section \ref{sec:4}. Finally section
\ref{sec:5} contains a concluding discussion of this manuscript.

\section{the wave functions of the charmed and bottomed baryons}
\label{sec:2}

 A baryon wave function that includes $qqqq\bar q$
components in addition to the conventional $qqq$ components may be
written in the following general form:

\begin{equation}
|B\rangle=\sqrt{P_{3q}}|qqq\rangle+\sqrt{P_{5q}}\sum_{i}A_{i}|qqqq_{i}
\bar{q_{i}}\rangle \, . \label{bwave}
\end{equation}
Here $P_{3q}$ and $P_{5q}$ are the probabilities of the $qqq$ and
$qqqq\bar{q}$ components respectively; the sum over i runs over all
the possible $qqqq_{i}\bar{q_{i}}$ components, and $A_{i}$ denotes
the amplitude of the corresponding 5q component. The flavor (and
spin) wave functions of every $qqqq_{i}\bar{q_{i}}$ is constructed
along with a calculation of the corresponding amplitudes $A_{i}$.

The wave functions of the $qqq$ components in the baryons are the
conventional ones, which are formed as combinations of the color,
space, flavor and spin wave functions with appropriate
Clebsch-Gordan coefficients. The ground states of the $C=+1$ baryons
can be in the spin states $[21]_{S}$ and $[3]_{S}$, i.e. the
flavor-spin states are $[3]_{FS}[21]_{F}[21]_{S}$ and
$[3]_{FS}[3]_{F}[3]_{S}$ in the static $qqq$ model, respectively.
Here we consider the previous ones, $J^{p}=1/2^{+}$, $\Sigma_{c}$,
$\Lambda_{c}$ and $\Xi_{c}^{a}$ baryons. The spectra of the $J=1/2$
charmed baryons, each with one charm quark, in the classic $qqq$
quark model, is just like that of the ones which have one or more
strange quarks in the octet. For instance, the $\Lambda_{c}$ and
$\Sigma_{c}$ spectra ought to look much like that of the $\Lambda$
and $\Sigma$, since a $\Lambda_{c}$ or a $\Sigma_{c}$ is obtained
from a $\Lambda$ or a $\Sigma$ by changing the s quark to a c quark.
But the $\Xi$ baryons have more than one strange quarks, then the
$\Xi_{c}$ spectra is some more complicated than the $\Xi$ spectra.
One may group the $\Xi_{c}$ baryons as that the strange and the
light quarks are symmetric or the strange and the light quark are
anti-symmetric, i.e. $\Xi_{c}^{s}$ and $\Xi_{c}^{a}$. We only
consider the $\Xi_{c}^{a}$ baryons in this manuscript.

As we have discussed above, the wave functions of the $qqq$
components in the charmed baryons are the conventional ones, with
flavor-spin symmetry $[3]_{FS}[21]_{F}[21]_{S}$, where $[f]_{i}$
denotes the Young Pattern with f being the sequence of the integers
that indicates the number of boxes in the successive rows of the
Young patterns. And this should be combined with the totally
symmetric orbital state $[3]_{X}$ and the antisymmetric color state
$[111]_{C}$ to form the completely antisymmetric state which is
required by the Pauli exclusion principle.

If the hyperfine interaction between the quarks depends on spin and
flavor are employed \cite{hyperfine}, the $qqqq$ subsystem with the
mixed spatial symmetry $[31]_{X}$ are expected to be the
configurations with the lowest energy, and therefore most likely to
form appreciable $qqqq\bar{q}$ components of the charm baryons.
Consequently the flavor-spin state of the $qqqq$ subsystem is most
likely totally symmetric: $[4]_{FS}$. Moreover in the case of the
$C=+1, J=1/2$ charmed baryons, the flavor-spin configurations of the
four quark subsystem $[4]_{FS}[22]_{F}[22]_{S}$, with one quark in
its first orbitally excited state, and the anti-quark in its ground
state, is likely to have the lowest energy \cite{hr}. The flavor
wave functions of the $qqqq\bar{q}$ components in the charmed
baryons are listed in Table \ref{table1}. The weight method
\cite{ma} has been employed here for the explicit construction of
the flavor wave functions in $qqqq_{i}\bar{q_{i}}$ configurations.

Actually, these flavor wave functions can be obtained directly from
that in Table \ref{table3} in Ref.\cite{bomm}, with the substitution
of the charm quark for the valence strange quark. Notice that there
is a typo in Table \textrm{3} in Ref. \cite{bomm}: the flavor wave
function in the fifth row with flavor symmetry $[22]_{F}$ should be:
$-(\sqrt{\frac{1}{3}}|ddsu\bar{u}\rangle-\sqrt{\frac{2}{3}}|ddss\bar{s}\rangle)$
. The new feature is the appearance of the states $\Xi^a_{c}$ and
$\Xi_{c}^s$, here we give the flavor wave functions of $\Xi_{c}^{a}$
baryons employing the weight diagram method, where the superscripts
'a' on the $\Xi_{c}$ states indicate that the valence light and
strange quarks are in antisymmetric states, the results are shown in
Table \ref{table1}.

\begin{table}[ht]\caption{\footnotesize The flavor wave functions of
the $qqqq\bar{q}$ components in the charmed baryons. \vspace{0.5cm}}
\footnotesize
\begin{tabular}{ll}

\hline

Baryons &  wave functions \\\hline

$\Lambda^ +_
c$&$-\sqrt{\frac{1}{2}}(|[udcu]_{[22]_{F}}\otimes\bar{u}\rangle
+|[udcd]_{[22]_{F}}\otimes\bar{d}\rangle)$\\

$\Sigma^{++}_c$&$\sqrt{\frac{1}{3}}|[uucd]_{[22]_{F}}\otimes\bar{d}\rangle
+\sqrt{\frac{2}{3}}|[uucs]_{[22]_{F}}\otimes\bar{s}\rangle$ \\

$\Sigma^{+}_{c}$&$-(\sqrt{\frac{1}{6}}|[udcu]_{[22]_{F}}\otimes\bar{u}\rangle
-\sqrt{\frac{1}{6}}|[udcd]_{[22]_{F}}\otimes\bar{d}\rangle)
+\sqrt{\frac{2}{3}}|[udcs]_{[22]_{F}}\otimes\bar{s}\rangle$\\

$\Sigma^{0}_{c}$&$-(\sqrt{\frac{1}{3}}|[ddcu]_{[22]_{F}}\otimes\bar{u}\rangle
-\sqrt{\frac{2}{3}}|[ddcs]_{[22]_{F}}\otimes\bar{s}\rangle)$ \\

$\Xi^{a+}_{c}$&$-\sqrt{\frac{1}{3}}|[uscu]_{[22]_{F}}\otimes\bar{u}\rangle
+\sqrt{\frac{1}{3}}|[uscd]_{[22]_{F}}\otimes\bar{d}\rangle
-\sqrt{\frac{1}{3}}|[uscs]_{[22]_{F}}\otimes\bar{s}\rangle$\\

$\Xi^{a0}_{c}$&$-\sqrt{\frac{1}{3}}|[dscu]_{[22]_{F}}\otimes\bar{u}\rangle
+\sqrt{\frac{1}{3}}|[dscd]_{[22]_{F}}\otimes\bar{d}\rangle)
-\sqrt{\frac{1}{3}}|[dscs]_{[22]_{F}}\otimes\bar{s}\rangle$\\

\hline

\end{tabular}
\label{table1}
\end{table}

In the case of the bottomed baryons, $\Sigma_{b}$ and
$\Sigma_{b}^{*}$, we can construct the flavor wave functions by the
similar way. For the $\Sigma_{b}$ baryons,
$J^{(p)}=\frac{1}{2}^{(+)}$, the flavor-spin configurations of the
$qqq$ components have the symmetry $[3]_{FS}[21]_{F}[21]_{S}$, while
that of the $\Sigma_{b}^{*}$, $J^{P}=\frac{3}{2}^{+}$, are
$[3]_{FS}[3]_{F}[3]_{S}$, and of course the orbital and color states
of all these baryons have the totally symmetric $[3]_{X}$ and
anti-symmetric $[111]_{C}$ configurations, respectively. For the
$qqqq\bar{q}$ components in these baryons, the $qqqq$ subsystem may
have the flavor-spin symmetry $[4]_{FS}[22]_{F}[22]_{S}$ and
$[4]_{FS}[31]_{F}[31]_{S}$, respectively, which are expected to have
the lowest energy \cite{hr}, therefore most likely to form notable
$qqqq\bar{q}$ components in the baryons. Then we can construct the
flavor wave functions of the $qqqq\bar{q}$ components in the
$\Sigma_{b}$ and $\Sigma_{b}^{*}$ baryons by employing the weight
method\cite{ma} or some substitutions of the beauty quark for the
valence strange quark in the flavor wave functions of the $\Sigma$
and $\Sigma^{*}$ baryons in Table \ref{table3} and Table
\ref{table2} in Ref.\cite{bomm}, respectively. The results are shown
in Table \ref{table2}.

\begin{table}[ht]\caption{\footnotesize The flavor wave functions of
the $qqqq\bar{q}$ components in the $\Sigma_{b}$ and
$\Sigma_{b}^{*}$ baryons . \vspace{0.5cm}} \footnotesize
\begin{tabular}{ll}

\hline

Baryons &   wave functions\\\hline

$\Sigma_{b}^{+}$&$\sqrt{\frac{1}{3}}|[uubd]_{[22]_{F}}\otimes\bar{d}\rangle
+\sqrt{\frac{2}{3}}|[uubs]_{[22]_{F}}\otimes\bar{s}\rangle$ \\

$\Sigma_{b}^{-}$&$-(\sqrt{\frac{1}{3}}|[ddbu]_{[22]_{F}}\otimes\bar{u}\rangle
-\sqrt{\frac{2}{3}}|[ddbs]_{[22]_{F}}\otimes\bar{s}\rangle)$ \\

$\Sigma_{b}^{*+}$&$\sqrt{\frac{1}{6}}|[uubu]_{[31]_{F}}\otimes\bar{u}\rangle
-\sqrt{\frac{1}{2}}|[uubd]_{[31]_{F}}\otimes\bar{d}\rangle
-\sqrt{\frac{1}{3}}|[uubs]_{[31]_{F}}\otimes\bar{s}\rangle$ \\

$\Sigma_{b}^{*-}$&$\sqrt{\frac{1}{2}}|[ddbu]_{[31]_{F}}\otimes\bar{u}\rangle
+\sqrt{\frac{1}{6}}|[ddbd]_{[31]_{F}}\otimes\bar{d}\rangle
-\sqrt{\frac{1}{3}}|[ddbs]_{[31]_{F}}\otimes\bar{s}\rangle$ \\

\hline

\end{tabular}
\label{table2}
\end{table}

 Here the flavor-spin decomposition of the states with the mixed
symmetry $[4]_{FS}[22]_{F}[22]_{S}$ and $[4]_{FS}[31]_{F}[31]_{S}$
are:
\begin{eqnarray}
|[4]_{FS}[22]_F[22]_S \rangle &=& {1\over
\sqrt{2}}\{[22]_{F_1}[22]_{S_1} +[22]_{F_2}[22]_{S_2}\}\, ,
\label{lowesta}\\
|[4]_{FS}[31]_F[31]_S \rangle &=& {1\over
\sqrt{3}}\{[31]_{F_1}[31]_{S_1}
+[31]_{F_2}[31]_{S_2}+[31]_{F_3}[31]_{S_3}\} .
\end{eqnarray}
And the notation of $[4]_{FS}[22]_{F}[22]_{S}$ is a shorthand for
the Young tableaux decomposition:
\begin{eqnarray}
&&[4]_{FS}[22]_F[22]_S\,:\,\Yvcentermath1
{\yng(4)}_{FS}\,{\yng(2,2)}_F\,{\yng(2,2)}_S \,,
\end{eqnarray}
and the notation of the $[4]_{FS}[31]_{F}[31]_{S}$ is:
\begin{eqnarray}
&&[4]_{FS}[31]_F[31]_S\,:\,\Yvcentermath1
{\yng(4)}_{FS}\,{\yng(3,1)}_F\,{\yng(3,1)}_S \,.
\end{eqnarray}

Above the wave functions of the charmed and bottomed baryons have
been given. Furthermore, the explicit forms of the flavor and spin
configurations $[22]_{F}$, $[31]_{F}$, $[22]_{S}$ and $[31]_{S}$ can
be found in Ref. \cite{zou2}. The orbital and color states of the
four quarks subsystem of all the $qqqq\bar{q}$ components are the
mixed symmetry $[31]_{X}$ and $[211]_{c}$, respectively, which are
restricted by the Pauli exclusive principle to form a completely
anti-symmetric state. The explicit form of the $[31]$ and $[211]$
configurations can be found in Ref. \cite{chen}. For instance, we
give the explicit wave function of the $uucd\bar{d}$ component in
the $\Sigma^{++}_{c}$ baryon.

\section{the magnetic moments of the charmed baryons}
\label{sec:3}

 In the non-relativistic quark model, the magnetic moment of a baryon is
defined as the expectation value of the following magnetic moment
operator in the corresponding baryon state (in units of nuclear
magnetons):

\begin{equation}
\hat{\mu}=\sum_{i}\frac{Q_{i}M_{N}}{em_{i}}(\hat{l}_{iz}+\hat{\sigma}_{iz}).
\label{op}
\end{equation}
Here the sum over i runs over the quark content of the baryon, and
$Q_{i}$ denotes the corresponding electric charge of the $ith$ quark
and $m_{i}$ is the constituent quark mass, e is the electric charge
of the proton.

On the other side, with the combination of $qqq$ and $qqqq\bar q$
state wave functions (\ref{bwave}) the magnetic moment will have
contributions from the diagonal matrix elements of the operator
(\ref{op}) between the $qqq$ component and the $qqqq\bar q$
components, respectively, and also from the off-diagonal terms, i.e.
the transition matrix elements between the $qqq$ and the $qqqq\bar
q$ components.  These will depend both on the explicit wave function
model and the model for the $q\bar{q}-\gamma$ vertices. If these
vertices are taken to have the elementary forms $\bar{v}(p')\gamma
u(p)$ and no account is taken of the interaction between the
annihilating $q\bar{q}$ pair and the baryon, in the non-relativistic
approximation, the transition operator which involve $q\bar q$ pair
annihilation and creation, may take form in momentum space as
follow:
\begin{eqnarray}
\hat{T}&=&Q_{i}\langle\vec{p}^{'}_{i}|\vec{j}_{i}
|\vec{p}_{i}\rangle_{z(anni)}\,,\nonumber\\
&=&\sum_{i}Q_{i}\hat{\sigma}_{i}\, . \label{ndop}
\end{eqnarray}
Here the sum over i runs over all the possible $qqqq_{i}\bar{q_{i}}$
components, and $Q_{i}$ denotes the electric charge of the
corresponding quark $q_{i}$. Note that the annihilation matrix
elements of the current operator above just contribute to the
non-diagonal term of the magnetic form factor, which will contribute
to the magnetic moments in the $q^{2}=0$ limit, for instance, we
give the calculations of the transition element between the $uuc$
and $uucd\bar{d}$ components in the $\Sigma^{++}_{c}$ baryons in
appendix \ref{sec:c}.

The calculations of these non-diagonal contributions to the magnetic
moments call for a specific orbital wave functions model. Here, for
simplicity, harmonic oscillator constituent quark model wave
functions are employed:

\begin{eqnarray}
\varphi_{00}(\vec p;\omega)&=&{1\over (\omega^2\pi)^{3/4}}
\exp\{-\frac{p^{2}}{2\omega^{2}}\}\, ,\ \\
\varphi_{1m}(\vec p;\omega)&=&\sqrt{2}\,{p_{m}\over \omega}\,
\varphi_{00}(\vec p;\omega)\, .
\end{eqnarray}
Here $\varphi_{00}(\vec{p};\omega)$ and
$\varphi_{1,m}(\vec{p};\omega)$ are the s-wave and p-wave orbital
wave functions of the constituent quarks, respectively. The
oscillator parameters of the $qqq$ and $qqqq\bar{q}$ components,
$\omega_{3}$ and $\omega_{5}$, will in general be different. The
parameter $\omega_3$ may be determined by the baryon radius as
$\omega_3=1/\sqrt{<r^2>}$. Here we employ the value
$\omega_{3}\sim250$ MeV. The parameter $\omega_5$ may be treated as
a free phenomenological parameter. In ref.\cite{zou3} it was noted
that the best description of the extant empirical strangeness form
factors is obtained with $\omega_5\sim$ 1 GeV, which would imply
that the $qqqq\bar q$ component is very compact.

With the wave functions and the magnetic moment operators above, the
charmed baryons magnetic moments can be expressed as the sum of the
diagonal matrix elements in the $qqq$ and $qqqq\bar{q}$ subspaces
and the off-diagonal transition matrix elements  of the form
$qqq\rightarrow qqqq\bar{q}$ and $qqqq\bar{q}\rightarrow qqq$. The
former only depend on the group theoretical factors, while the
latter also depend on the spatial wave function model. The diagonal
contributions to the charmed baryons magnetic moments are listed in
table \ref{table3}.

\begin{table}[ht]\caption{\footnotesize The diagonal contributions
to the magnetic moments of the charmed baryons. \vspace{0.5cm}}
\footnotesize
\begin{tabular}{ll}

\hline

Baryons &  magnetic moments \\\hline

$\Lambda^ +_ c$&$P_{3q}\frac{2M_{N}}{3m_{c}}+
P_{(\Lambda_{c}^{+})u\bar{u}}(\frac{7M_{N}}{18m}+
\frac{M_{N}}{9m_{c}})-P_{(\Lambda_{c}^{+})d\bar{d}}
(\frac{M_{N}}{9m}-\frac{M_{N}}{9m_{c}})$\\

$\Sigma^{++}_c$&$P_{3q}(\frac{8M_{N}}{9m}-
\frac{2M_{N}}{9m_{c}})+P_{(\Sigma^{++}_{c})d\bar{d}}
(\frac{M_{N}}{18m}+\frac{M_{N}}{9m_{c}})
+P_{(\Sigma^{++}_{c})s\bar{s}}(\frac{2M_{N}}{9m}-
\frac{M_{N}}{6m_{s}}
+\frac{M_{N}}{9m_{c}})$ \\

$\Sigma^{+}_{c}$&$P_{3q}(\frac{2M_{N}}{9m}-\frac{2M_{N}}{9m_{c}}
)+P_{(\Sigma^{+}_{c})u\bar{u}}
(\frac{7M_{N}}{18m}+\frac{M_{N}}{9m_{c}})-
P_{(\Sigma^{+}_{c})d\bar{d}}(\frac{M_{N}}{9m}- \frac{M_{N}}
{9m_{c}})+P_{(\Sigma^{+}_{c})s\bar{s}}
$\\

&$(\frac{M_{N}}{18m}-\frac{M_{N}}{6m_{s}}+\frac{M_{N}}{9m_{c}})$\\

$\Sigma^{0}_{c}$&$-P_{3q}(\frac{4M_{N}}{9m}+\frac{2M_{N}}{9m_{c}})
+P_{(\Sigma^{0}_{c})u\bar{u}}
(\frac{2M_{N}}{9m}+\frac{M_{N}}{9m_{c}})-
P_{(\Sigma^{0}_{c})s\bar{s}}(\frac{M_{N}}{9m}+
\frac{M_{N}}{6m_{s}}-\frac{M_{N}}{9m_{c}})$ \\

$\Xi^{a+}_{c}$&$P_{3q}\frac{2M_{N}}{3m_{c}}+
P_{(\Xi^{a+}_{c})u\bar{u}}(\frac{4M_{N}}{9m}-
\frac{M_{N}}{18m_{s}}+\frac{M_{N}}{9m_{c}})+
P_{(\Xi^{a+}_{c})d\bar{d}}(-\frac{M_{N}}{18m}-
\frac{M_{N}}{18m_{s}}+\frac{M_{N}}{9m_{c}})$\\
&$+P_{(\Xi^{a+}_{c})s\bar{s}}(\frac{M_{N}}{9m}-
\frac{2M_{N}}{9m_{s}}+\frac{M_{N}}{9m_{c}})$\\

$\Xi^{a0}_{c}$&$ P_{3q}\frac{2M_{N}}{3m_{c}}+
P_{(\Xi^{a0}_{c})u\bar{u}}(\frac{5M_{N}}{18m}-
\frac{M_{N}}{18m_{s}}+\frac{M_{N}}{9m_{c}})+
P_{(\Xi^{a0}_{c})d\bar{d}}(-\frac{2M_{N}}{9m}-
\frac{M_{N}}{18m_{s}}+\frac{M_{N}}{9m_{c}})$\\
&$+P_{(\Xi^{a0}_{c})s\bar{s}}(-\frac{M_{N}}{18m}-
\frac{2M_{N}}{9m_{s}}+\frac{M_{N}}{9m_{c}})$\\

$\Sigma^{+}_{c}\rightarrow\Lambda_{c}^{+}$&
$-P_{3q}\frac{M_{N}}{\sqrt{3}m}
+\frac{1}{4\sqrt{3}}P_{5q}\frac{M_{N}}{m}$\\

\hline

\end{tabular}
\label{table3}
\end{table}

Here the factors $P_{(B)q_i\bar{q_i}}$ are the probabilities of the
$qqqq_i\bar{q}_i$ components in the baryon $B$. These are related to
the corresponding amplitudes $A_i$ and the probability of the
$qqqq\bar{q}$ components (\ref{bwave}) as:
\begin{equation}
P_{(B)q_{i}\bar{q}_{i}}=P_{5q}A_{i}^{2}\,.\
\end{equation}

The contributions of the non-diagonal matrix elements are listed in
table \ref{table4}.

\begin{table}[ht]\caption{\footnotesize The off-diagonal contributions
to the magnetic moments of the charmed baryons. \vspace{0.5cm}}
\footnotesize
\begin{tabular}{ll}

\hline

Baryons &  magnetic moments \\\hline

$\Lambda^ +_ c$&$-\frac{1}{3}F_{35}(P_{5q})$\\

$\Sigma^{++}_c$&$-\frac{2\sqrt{3}}{9}F_{35}
(P_{(\Sigma^{++}_{c})d\bar{d}})
-\frac{2\sqrt{3}}{9}F_{35}(P_{(\Sigma^{++}_{c})s\bar{s}})$ \\

$\Sigma^{+}_{c}$&$\frac{2\sqrt{6}}{9}F_{35}
(P_{(\Sigma^{+}_{c})u\bar{u}})-
\frac{\sqrt{6}}{9}F_{35}(P_{(\Sigma^{+}_{c})d\bar{d}})
-\frac{2\sqrt{3}}{9}F_{35} (P_{(\Sigma^{+}_{c})s\bar{s}})
$\\

$\Sigma^{0}_{c}$&$\frac{4\sqrt{3}}{9}F_{35}(P_{(\Sigma^{0}_{c})u\bar{u}})
-\frac{2\sqrt{3}}{9}F_{35}
(P_{(\Sigma^{0}_{c})s\bar{s}})$ \\

$\Xi^{a+}_{c}$&$\frac{2\sqrt{2}}{3}F_{35}
(P_{(\Xi^{a+}_{c})u\bar{u}})+\frac{1}{3}F_{35}
(P_{(\Xi^{a+})d\bar{d}}) -\frac{\sqrt{2}}{3}F_{35}
(P_{(\Xi^{a+})s\bar{s}})$\\

$\Xi^{a0}_{c}$&$\frac{2}{3}F_{35}(P_{(\Xi^{a0}_{c})u\bar{u}})+\frac{\sqrt{2}}{3}F_{35}
(P_{(\Xi^{a0}_{c})d\bar{d}}) -\frac{\sqrt{2}}{3}F_{35}
(P_{(\Xi^{a0}_{c})s\bar{s}})$\\

$\Sigma^{+}_{c}\rightarrow\Lambda_{c}^{+}$&
$-\frac{2\sqrt{3}}{3}F_{35}(P_{5q})$\\

\hline

\end{tabular}
\label{table4}
\end{table}

The functions $F_{35}(P_{(B)q\bar{q}})$ in table \ref{table4} are
defined as
\begin{equation}
F_{35}(P_{(B)q\bar{q}})=C_{35}\frac{M_{N}}{\omega_{5}}
\sqrt{P_{3q}P_{(B)q\bar{q}}}\, ,
\end{equation}
where the factor $C_{35}$,
\begin{equation}
C_{35}=(\frac{2\omega_{3}\omega_{5}}{\omega_{3}^{2}+
\omega_{5}^{2}})^{9/2}\, , \label{c35}
\end{equation}
is the overlap integral of the s-wave wave functions of the quarks
in the $qqq$ and $qqqq\bar{q}$ configurations.

Note that here we have taken all the phase factors for the
non-diagonal matrix element between the $qqq$ and $qqqq\bar{q}$
components to be +1.

As in the above expressions, we can divided the 5q contributions
into the oscillator model independent table \ref{table3} and
dependent terms table \ref{table4}. Furthermore, we can also extract
the hidden light and strange flavors contributions to the charmed
baryons magnetic moments in the previous expressions,respectively.
We can find that in table \ref{table3} the hidden flavor
contributions to the charmed baryon magnetic moments are
proportional to $P_{(B)q_i\bar{q_i}}$ and inversely proportional to
the constituent quark mass, and in table \ref{table4} they are
proportional to the factor $A_{3q}A_{(B)q_i\bar{q_i}}$, which are
the amplitudes of the $qqq$ and $qqqq\bar{q}$ components
respectively. From Ref.\cite{zou1,zou2,zou3,bomm}, we know that the
possibilities of the $qqqq\bar{q}$ components may be $10-40\%$,
therefore, the hidden flavor contributions to the charmed baryons
should be significant.

To obtain the numerical results of the charmed magnetic moments,
here the parameters are the masses of the constituent quarks, the
oscillator model parameter $\omega_{5}$, and the probability of the
$qqqq\bar{q}$ component. We set the constituent quark masses to be
$m_{u}=m_{d}=m=280$ MeV, $m_{s}=420$ MeV and $m_{c}=1600$ MeV,
respectively. The oscillator parameter $\omega_{5}$ is taken to have
the value 0.57GeV and 1GeV respectively, which corresponds to a
compact extension of the $qqqq\bar{q}$. The value 0.57 GeV may be
the best value for explaining the magnetic moments of the baryon
octet, which has been shown in Ref. \cite{bomm}. The probability of
the $qqqq\bar{q}$ components are taken as the tentative value
$P_{5q}=20\%$. The numerical results are shown in Table
\ref{table5}, comparing with the results from the conventional
static quark model, the chiral perturbation theory and the values
predicted on the basis of the bound state soliton model.

\begin{table}[h]\caption{\footnotesize Magnetic moments of
the charmed baryons, the results are in units of nuclear magnetons.
The column $qqq$ contains the results of the conventional quark
model from Refs. \cite{riska} and column C the results from the
chiral perturbation theory \cite{perturbation}, in column S the
values predicted in Ref. \cite{soliton} on the basis of the bound
state soliton model are given. Column $P_{ls}$ are the present
results from the light and strange quarks contributions, which are
the diagonal matrix elements, and $P_{c}$ the charm quark
contributions. The non-diagonal contributions are listed in column
$P_{n1}$ and $P_{n2}$ with $\omega_{5}=0.57$ GeV and $\omega_{5}=1$
GeV, respectively. Finally, the present results are listed in column
$P_{1}$ with $\omega_{5}=0.57$ GeV and $P_{2}$ with $\omega_{5}=1$
GeV. } \vspace{0.5cm} \footnotesize

\begin{tabular}{ccccccccccc}

\hline

Baryons           & $qqq$  & $P_{ls}$ & $P_{c}$& $P_{n1}$ &
$P_{n2}$& $P_{1}$ & $P_{2}$ & & C & S
\\\hline

$\Lambda_{c}^{+}$ &  0.38 &0.09&0.33&-0.06&-0.004& 0.36  & 0.41 &     &  0.37 &   0.28 \\

$\Xi_{c}^{a+}$    &  0.38 &0.06&0.33&-0.08&-0.006& 0.46  & 0.39 &     &  0.42 &   0.28 \\

$\Xi_{c}^{a0}$    &  0.38 &-0.05&0.33&-0.06&-0.005& 0.34  & 0.28 &     &  0.32 &   0.28 \\

$\Sigma_{c}^{++}$ &  2.33 &2.44&-0.09&-0.09&-0.007& 2.26  & 2.35 &     &       &   2.76 \\

$\Sigma_{c}^{+}$  &  0.49 &0.60&-0.09&-0.03&-0.003& 0.48  & 0.51 &     &       &   0.59 \\

$\Sigma_{c}^{0}$  & -1.35 &-1.24&-0.09&-0.02&0.002& -1.31 &-1.33 &     &       &  -1.35 \\

$\Sigma_{c}^+\rightarrow \Lambda_{c}^{+}$  & -1.59 &-1.45&0&-0.19&-0.014& -1.64 & -1.47 &  &  \\

\hline

\end{tabular}
\label{table5}
\end{table}

As we can see in column $P_{1}$ in table \ref{table5}, the magnetic
moments of the $\Xi_{c}^{a+(0)}$ are different from that of the
$\Lambda_{c}^{+}$ by including the $qqqq\bar{q}$ components
contributions. Especially for that of the $\Xi_{c}^{a+}$ baryon,
which differ from that of the $\Lambda_{c}^{+}$ baryon by about 28
\% . It's interesting that the average value of the $\Xi_{c}^{a+}$,
$\Xi_{c}^{a0}$ and $\Lambda_{c}^{+}$ baryons magnetic moments is
about $0.39\mu_{N}$, it's consistent with the value $0.38\mu_{N}$ of
the classic static quark model, and it's also in excellent agreement
with the chiral perturbation theory value
$0.37\mu_{N}$\cite{perturbation}. In Ref.\cite{perturbation}, the
leading long-distance contributions to the magnetic moments of the
$\Lambda_{c}^{+}$, $\Xi_{c}^{a+}$ and $\Xi_{c}^{a0}$ charmed baryons
from the spin symmetry breaking $\Sigma^{*}_{c}-\Sigma_{c}$ mass
splitting in chiral perturbation theory are computed, these are
nonanalytic in the pion mass and arise from calculable one-loop
graphs. From that, the differences of the $\Lambda_{c}^{+}$,
$\Xi_{c}^{a+}$ and $\Xi_{c}^{a0}$ magnetic moments are independent
of the charm quark mass. As shown in table \ref{table3} and
\ref{table4}, the charm quarks contributions from the diagonal terms
to the $\Lambda_{c}^{+}$, $\Xi_{c}^{+}$ and $\Xi_{c}^{0}$ magnetic
moments are same in our results, which means that the difference of
these magnetic moments only comes from the contributions of the
light and strange quark. Consequently, the differences of these
magnetic moments are also independent of the charm quark mass for
that all the off-diagonal terms are quark masses independent. The
similar conclusion may come from that both of the two method are
based on the contributions of the sea quark. In column $S$, the
results of Ref. \cite{soliton} are given, which are similar to that
of the non-relativistic constituent $qqq$ quark model.

The magnetic moment of $\Lambda_{c}^{+}$ is not very sensitive to
the proportion of the $qqqq\bar{q}$ components, the others vary a
bit with $P_{5q}$ changed. For instance, if $P_{5q}$ increase to the
value $P_{5q}=0.50$, the parameter $\omega_{5}$ is set to be $0.57$
GeV , these magnetic moments will be $\mu_{\Lambda_{c}}=0.35$ n.m. ,
$\mu_{\Xi_{c}^{a+}}=0.54$ n. m.  and $\mu_{\Xi_{c}^{a0}}=0.23$ n.
m..

These magnetic moments have also been calculated in the relativistic
quark model in Ref. \cite{relativistic}, a bit corrections to the
results of the non-relativistic constituent quark model have been
given. As we can see in Ref. \cite{relativistic}, a bit difference
for the magnetic moments of the $\Lambda_{c}^{+}$, $\Xi_{c}^{a+}$
and $\Xi_{c}^{a0}$ can be also obtained in the relativistic quark
model. The bound state approach is employed to calculate these
magnetic moments in Ref. \cite{bound}. In that paper, they give a
smaller value for the $\Lambda_{c}$ baryon,
$\mu_{\Lambda_{c}^{+}}=0.12$ n. m..

\section{the magnetic moments of the $\Sigma_{b}$ and $\Sigma_{b}^{*}$ baryons}
\label{sec:4}

In the case of the bottomed baryons $\Sigma_{b}$, the lowest
$qqqq\bar{q}$ configuration are also expected to be that the $qqqq$
subsystem is assumed to have $[4]_{FS}[22]_{F}[22]_{S}$ mixed
symmetry, the flavor wave functions are shown in Table \ref{table2}.
The matrix elements of the magnetic moment operator (\ref{op}) in
these wave functions, and the corresponding $qqq$ wave functions
lead to the following diagonal contributions to the magnetic moments
are listed in table \ref{table6}.

\begin{table}[ht]\caption{\footnotesize The diagonal contributions
of the magnetic moments of the bottomed moments. \vspace{0.5cm}}
\footnotesize
\begin{tabular}{ll}

\hline

Baryons &   magnetic moments\\\hline

$\Sigma_{b}^{+}$&$P_{3q}(\frac{8M_{N}}{9m}+
\frac{M_{N}}{9m_{b}})+P_{(\Sigma^{+}_{b})d\bar{d}}
(\frac{M_{N}}{18m}-\frac{M_{N}}{18m_{b}})
+P_{(\Sigma^{+}_{b})s\bar{s}}(\frac{2M_{N}}{9m}-
\frac{M_{N}}{6m_{s}}-\frac{M_{N}}{18m_{b}})$ \\

$\Sigma_{b}^{-}$&$-P_{3q}(\frac{4M_{N}}{9m}-\frac{M_{N}}{9m_{b}})
+P_{(\Sigma^{-}_{b})u\bar{u}}
(\frac{2M_{N}}{9m}-\frac{M_{N}}{18m_{b}})-
P_{(\Sigma^{-}_{b})s\bar{s}}(\frac{M_{N}}{9m}+
\frac{M_{N}}{6m_{s}}+\frac{M_{N}}{18m_{b}})$ \\

$\Sigma_{b}^{*+}$&$P_{3q}(\frac{4M_{N}}{3m}-\frac{M_{N}}{3m_{b}})
+P_{(\Sigma_{b}^{*+})u\bar{u}}
(\frac{5M_{N}}{12m}+\frac{M_{N}}{24m_{b}})+P_{(\Sigma_{b}^{*+})d\bar{d}}
(\frac{7M_{N}}{8m}+\frac{M_{N}}{24m_{b}})$\\
&$+P_{(\Sigma_{b}^{*+})s\bar{s}}
(\frac{13M_{N}}{18m}+\frac{11M_{N}}{72m_{s}}+\frac{M_{N}}{24m_{b}})$ \\

$\Sigma_{b}^{*-}$&$P_{3q}(-\frac{2M_{N}}{3m}-\frac{M_{N}}{3m_{b}})
+P_{(\Sigma_{b}^{*-})u\bar{u}}
(-\frac{2M_{N}}{3m}+\frac{M_{N}}{24m_{b}})+P_{(\Sigma_{b}^{*-})d\bar{d}}
(-\frac{5M_{N}}{24m}+\frac{M_{N}}{24m_{b}})$\\
&$+P_{(\Sigma_{b}^{*-})s\bar{s}}
(-\frac{13M_{N}}{36m}+\frac{11M_{N}}{72m_{s}}+\frac{M_{N}}{24m_{b}})$ \\

\hline

\end{tabular}
\label{table6}
\end{table}

And the non-diagonal matrix elements are listed in table
\ref{table7}.

\begin{table}[ht]\caption{\footnotesize The off-diagonal contributions
of the magnetic moments of the bottomed moments. \vspace{0.5cm}}
\footnotesize
\begin{tabular}{ll}

\hline

Baryons &   magnetic moments\\\hline

$\Sigma_{b}^{+}$&$-\frac{2\sqrt{3}}{9}F_{35}
(P_{(\Sigma^{+}_{b})d\bar{d}})
-\frac{2\sqrt{3}}{9}F_{35}(P_{(\Sigma^{+}_{b})s\bar{s}})$ \\

$\Sigma_{b}^{-}$&$\frac{4\sqrt{3}}{9}F_{35}(P_{(\Sigma^{-}_{b})u\bar{u}})
-\frac{2\sqrt{3}}{9}F_{35} (P_{(\Sigma^{-}_{b})s\bar{s}})$ \\

$\Sigma_{b}^{*+}$&$\frac{\sqrt{3}}{6}F_{35}(P_{(\Sigma_{b}^{*+})u\bar{u}})
+\frac{1}{12}F_{35}(P_{(\Sigma_{b}^{*+})d\bar{d}})
+\frac{1}{12}F_{35}(P_{(\Sigma_{b}^{*+})s\bar{s}})$\\

$\Sigma_{b}^{*-}$&$\frac{1}{6}F_{35}(P_{(\Sigma_{b}^{*-})u\bar{u}})
-\frac{\sqrt{3}}{12}F_{35}(P_{(\Sigma_{b}^{*-})d\bar{d}})
+\frac{1}{12}F_{35}(P_{(\Sigma_{b}^{*-})s\bar{s}})$ \\

\hline

\end{tabular}
\label{table7}
\end{table}

For the $\Sigma_{b}^{*}$ baryons, the lowest energy $qqqq\bar{q}$
configurations are expected to be that the $qqqq$ subsystem is
assumed to have the $[4]_{FS}[31]_{F}[31]_{S}$ mixed symmetry, the
flavor wave functions are also shown in Table \ref{table2}. And the
diagonal matrix elements are listed in table \ref{table6}, and the
non-diagonal elements are listed in table \ref{table7}.

Note that the $qqqq$ subsystem of the $qqqq\bar q$ components in the
$\Sigma_{b}^{*}$ baryons can have both total angular momentum $J=1$
and $J=2$, which are required by the spin 3/2 of the baryons. Here
have assumed that $J=1$. With 20\% proportion of the $qqqq\bar q$
components assumed in the $\Sigma_{b}$ and $\Sigma_{b}^{*}$ baryons,
the above expressions lead to the numerical results in Table
\ref{table8}, compared with the results from only the $qqq$
components. Here we set the constituent mass of the beauty quark to
be $m_{b}=5.0$ GeV. As it's shown in Table \ref{table8}, the
influence of the light and strange $q\bar{q}$ pairs is significant.

\begin{table}[h]\caption{\footnotesize Magnetic moments of
the $\Sigma_{b}$ and $\Sigma_{b}^{*}$ baryons, the results are in
units of the nuclear magnetons. Column $qqq$ contains the results of
that from only the $qqq$ components. Column $P_{ls}$ are the present
results from the light and strange quarks contributions, which are
the diagonal matrix elements, and $P_{b}$ the beauty quark
contributions. The non-diagonal contributions are listed in column
$P_{n1}$ and $P_{n2}$ with $\omega_{5}=0.57$ GeV and $\omega_{5}=1$
GeV, respectively. Finally, column A contains the results including
the contributions of the $qqqq\bar{q}$ components with
$\omega_{5}=0.57$ GeV and column B the values with $\omega_{5}=1$
GeV. Here we set the constituent mass of the beauty quark to be
$m_{b}=5.0$ GeV.} \vspace{0.5cm} \footnotesize
\begin{tabular}{cccccccc}

\hline

  baryons          & $qqq$  &$P_{ls}$&$P_{b}$&$P_{n1}$& $P_{n2}$ & $A$      &   B         \\\hline

$\Sigma_{b}^{+}$   & 3.00  &2.44&0.01&-0.09&  -0.007 &  2.37    &   2.45      \\

$\Sigma_{b}^{-}$   & -1.47 &-1.24&0.01&0.02& 0.002  &  -1.20   &  -1.22      \\

$\Sigma_{b}^{*+}$  & 4.40  &4.23&-0.05&0.05&  0.004 &  4.23    & 4.18       \\

$\Sigma_{b}^{*-}$  & -2.30  &-2.10&-0.05&0.01&  0.001&  -2.13   & -2.14       \\

\hline

\end{tabular}
\label{table8}
\end{table}

The numerical results will vary a bit with the proportion of the
$qqqq\bar{q}$ components changed. For instance, if we increase it to
the value $P_{5q}=0.50$, these magnetic moments will be
$\mu_{\Sigma_{b}^{+}}=1.54$ n. m. , $\mu_{\Sigma_{b}^{-}}=-0.84$ n.
m., $\mu_{\Sigma^{*+}_{b}}=3.91$ n. m. and
$\mu_{\Sigma^{*-}_{b}}=-1.90$ n. m., which means that the increasing
of the $P_{5q}$ will make the absolute value of these magnetic
moments decreased.

As in Ref. \cite{perturbation}, they give the prediction that the
magnetic moments of the b-baryons in the $\bar{3}$ of $SU(3)$ can be
obtained by the same analysis as the c-baryons in $\bar{3}$, which
means that some correction will be added to the value
$-\frac{1}{3}\frac{1}{m_{b}}$ from the constituent quark model ,
which will lead to three different magnetic moments. Here if we take
into account the contributions of the $qqqq\bar{q}$ components, we
can also get the same prediction, and the difference of the three
magnetic moments should be independent with constituent mass of the
bottom quark.

\section{conclusions}
\label{sec:5}

Here the charmed and bottomed baryons magnetic moments have been
computed within the framework of the non-relativistic constituent
quark model, and the influence of the light and strange $q\bar{q}$
pairs has been investigated. In addition explicit expressions of the
wave functions for all the possible $qqqq\bar q$ components in the
charmed and bottomed baryons are given. The calculated magnetic
moments include the contributions from the $qqq$ components and the
$qqqq\bar q$ components with both of light and strange $q\bar q$
pairs. The magnetic moment expressions readily allow separation of
the strangeness components from $s\bar s$ pairs as well as
individual components from $u\bar u$ and $d\bar d$ pairs.

Our main conclusion is that, the $qqqq\bar{q}$ components, i.e. the
hidden light and strange flavors, have notable contributions to the
charmed baryons magnetic moments. Of course, the additional
$qqqq\bar{q}$ components contributions also lead to notable
corrections to the bottomed baryons magnetic moments. The numerical
results are shown in Table \ref{table5} and \ref{table8}, with a
$\sim20\%$ admixtures of the $qqqq\bar{q}$ components in the
baryons.

As we have discussed in Ref. \cite {bomm}, the restriction in the
calculation of the charmed baryon and $\Sigma_{b}$ magnetic moments
to the configuration with flavor-spin symmetry
$[4]_{FS}[22]_F[22]_S$ given in Table \ref{table1} and \ref{table2},
is motivated by its expected low energy. The configuration with
$[4]_{FS}[31]_F[31]_S$ symmetry is expected to have the next lowest
energy and to give an at most very insignificant contribution to the
magnetic moments. In the case of the $\Sigma_{b}^{*}$ this
configuration is, however, expected to have the lowest energy as the
configuration $[4]_{FS}[22]_F[22]_S$ cannot contribute \cite{bomm}.

It is of course not in any way obvious that those $qqqq\bar q$
configurations that have the lowest energy for a given hyperfine
interaction model should have the largest probability in the
baryons. The main terms should be expected to be those with the
strongest coupling to the $qqq$ component. This coupling depends
both on the confining interaction and (inversely) on the difference
in energy from the rest energy of the baryons.

\begin{acknowledgments}

 The author is grateful to Prof. D. O. Riska for suggestion to calculate
the magnetic moments of the charmed and bottomed baryons and
valuable discussions, and Prof. B. S. Zou for patient guidance.
Research supported in part by the National Natural Science
Foundation of China under grants Nos.10225525 \& 10435080.

\end{acknowledgments}

\begin{appendix}

\section{the explicit wave functions of the $uuc$ and $uucd\bar{d}$ components in the
$\Sigma^{++}_{c}$ baryon}
\label{sec:a}

The wave function of the $uuc$ component is the traditional one, and
that of the $uucd\bar{d}$ component may be expressed as
\begin{equation}
\begin{array}{ll}|\Sigma_{c}^{++},+1/2\rangle_{5q}=
&\sqrt{\frac{1}{2}}\sum_{ab}\sum_{ms_{z}}C^{\frac{1}{2}\frac{1}{2}}
_{1m,\frac{1}{2}s_{z}}C^{[1^{4}]}_{[31]_{a} [211]_{a}}\\
&[31]_{X,m}(a) [22]_{F}(b)[22]_{S}(b)[211]_{C}(a)\bar\chi_{s_{z}}
\psi(\{\vec{p}_i\}) \, . \label{wfc}
\end{array}
\end{equation}
Here the color, space and flavor-spin wave functions of the $qqqq$
subsystem have been denoted by their Young patterns, respectively,
the sum over $a$ runs over the three configurations of the
$[211]_{C}$ and $[31]_{X}$ representations of $S_{4}$, and the sum
over $b$ runs over the two configurations of the $[22]$
representation of $S_{4}$ respectively, and $C^{[1^{4}]}_{[31]_{a}
[211]_{a}}$ denotes the $S_{4}$ Clebsch-Gordan coefficients of the
representations $[1111][31][211]$, $\bar\chi_{s_{z}}$ is the spin
state of the anti-quark.

The explicit color-space part of the wave function (\ref{wfc}) can
be expressed in the form
\begin{equation}
\psi_{C}(\{\vec{p}_i\})=\frac{1}{\sqrt{3}}\{[211]_{C3}[31]_{X1}
-[211]_{C2}[31]_{X2}+[211]_{C1}[31]_{X3}\}\varphi_{00}(\vec{p}_{5})
\label{CX}
\end{equation}
where the explicit forms of the $[31]_{Xi}$ are defined as
\cite{chen}
\begin{eqnarray}
|[31]_{X_{1}}>&=&\frac{1}{\sqrt{12}}[3|0001\rangle-|0010\rangle
-|0100\rangle-|1000\rangle]\, ,\label{combs1}\\
|[31]_{X_{2}}>&=&\frac{1}{\sqrt{6}}[2|0010\rangle-|0100\rangle-|1000\rangle]\, ,\label{combs2}\\
|[31]_{S_{3}}>&=&\frac{1}{\sqrt{2}}[|0100\rangle-|1000\rangle]\,.
\label{combs3}
\end{eqnarray}
Here the numbers $'0'$ and $'1'$ denote the ground and excited
orbital states of the corresponding quark, for instance,
\begin{equation}
|0001\rangle=\varphi_{00}(\vec{p}_{1})\varphi_{00}(\vec{p}_{2})\varphi_{00}(\vec{p}_{3})\varphi_{1m}(\vec{p}_{4})
\end{equation}

The flavor-spin configuration of the $uuc$ component in the
$\Sigma^{++}_{c}$ baryon has the symmetry
$[3]_{FS}[21]_{F}[21_{S}]$, which can be expressed as:
\begin{equation}
|[3]_{FS}[21]_{F}[21_{S}]\rangle=\frac{1}{\sqrt{2}}(|[21]_{F1}\rangle|[21]_{s1}\rangle
+|[21]_{F2}\rangle|[21]_{S2}\rangle) \label{qqq}
\end{equation}
The explicit forms of $[21]_{F1}$, $[21]_{s1}$, $[21]_{F2}$ and
$[21]_{S2}$ are:
\begin{eqnarray}
|[21]_{F1}\rangle \!&=&\sqrt{\frac{1}{6}}(|ucu\rangle+|cuu\rangle-2|uuc\rangle)\\
|[21]_{S1}\rangle
\!&=&\sqrt{\frac{1}{6}}(|\uparrow\downarrow\uparrow\rangle+|
\downarrow\uparrow\uparrow\rangle-2|\uparrow\uparrow\downarrow\rangle)\\
|[21_{F2}]\rangle \!&=&\sqrt{\frac{1}{2}}(|ucu\rangle-|cuu\rangle)\\
|[21_{S2}]\rangle
\!&=&\sqrt{\frac{1}{2}}(|\uparrow\downarrow\uparrow\rangle-
|\downarrow\uparrow\uparrow\rangle)
\end{eqnarray}

The flavor-spin configuration of the $uucd$ subsystem of the
$uucd\bar{d}$ component is $[4]_{FS}[22]_{F}[22_{S}]$, which can be
expressed as:
\begin{equation}
|[4]_{FS}[22]_{F}[22_{S}]\rangle=\frac{1}{\sqrt{2}}(|[22]_{F1}\rangle|[22]_{s1}\rangle+
|[22]_{F2}\rangle|[22]_{S2}\rangle)
\end{equation}
here the CG coefficient $\frac{1}{\sqrt{2}}$ is just the same one in
equation (\ref{wfc}). The explicit forms of $[22]_{F1}$,
$[22]_{s1}$, $[22]_{F2}$ and $[22]_{S2}$ are:
\begin{eqnarray}
|[22]_{F_{1}}\rangle \!&=&
\frac{1}{\sqrt{24}}[2|uudc\rangle+2|uucd\rangle+2|dcuu\rangle+2|cduu\rangle
-|duuc\rangle-|uduc\rangle\nonumber\\
&&-|cudu\rangle- |ucdu\rangle-|cuud\rangle-|ducu\rangle-|ucud\rangle
-|udcu\rangle]\\
|[22]_{S_{1}}\rangle \!&=&
\frac{1}{\sqrt{12}}[2|\uparrow\uparrow\downarrow\downarrow\rangle+
2|\downarrow\downarrow\uparrow\uparrow\rangle-|\downarrow\uparrow\uparrow\downarrow\rangle
-|\uparrow\downarrow\uparrow\downarrow\rangle
-|\downarrow\uparrow\downarrow\uparrow\rangle-|\uparrow\downarrow\downarrow\uparrow\rangle\\
|[22]_{F_{2}}\rangle \!&=&\!
\frac{1}{\sqrt{8}}[|uduc\rangle+|cudu\rangle+|ducu\rangle+|ucud\rangle-
|duuc\rangle-|ucdu\rangle\nonumber\\
&&-|cuud\rangle-|udcu\rangle]\\
 |[22]_{S_{2}}\rangle \!&=&
\frac{1}{2}[|\uparrow\downarrow\uparrow\downarrow\rangle+|\downarrow\uparrow\downarrow\uparrow\rangle
-|\downarrow\uparrow\uparrow\downarrow\rangle-|\uparrow\downarrow\downarrow\uparrow\rangle]
\label{FS}
\end{eqnarray}

\section{the diagonal contribution of the $uucd\bar{d}$ component}
\label{sec:b}
 Here we have given the explicit wave functions of the $uuc$
and $uucd\bar{d}$ components. In equation (\ref{op}), the sum runs
over all the quark and anti-quark content in the $uucd\bar{d}$
component. The four-quark subsystem has no contribution to the
expectation value of the operator $\sigma_{iz}$ since the total spin
is $S=0$ for the $[22]_{S}$ symmetry. Consequently, it only
contribute to the expectation value of the operator $\l_{iz}$. For
the fourth quark, the contribution is:
\begin{equation}
\frac{2}{3}<\Psi_C|l_{4z}|\Psi_C><[4]_{FS}|\frac{Q_{4}M_{N}}{em_{4}}|[4]_{FS}>
\end{equation}
Here the factor $\frac{2}{3}$ is the square of the Clebsch-Gordan
coefficient $C^{\frac{1}{2}\frac{1}{2}}
_{11,\frac{1}{2}-\frac{1}{2}}$, where -1/2 is the z-component of the
anti-quark spin. After some calculations, we can get
\begin{eqnarray}
<\Psi_C|l_{4z}|\Psi_C>&=&\frac{1}{4}\nonumber\\
<[4]_{FS}|\frac{Q_{4}M_{N}}{em_{4}}|[4]_{FS}>
&=&\frac{1}{4}(\frac{2M_{N}}{3m_{c}}+\frac{M_{N}}{m})
\end{eqnarray}
so the contributions of the four-quark subsystem should be:
\begin{equation}
\mu_{uucd}=\frac{1}{6}(\frac{2M_{N}}{3m_{c}}+\frac{M_{N}}{m})
\end{equation}
The contribution of the fifth quark, which refers to the anti-quark,
is
\begin{eqnarray}
\mu_{\bar{d}}&=&\frac{Q_{\bar{d}}M_{N}}{em}[(C^{\frac{1}{2}\frac{1}{2}}
_{11,\frac{1}{2}-\frac{1}{2}})^{2}
\langle\frac{1}{2},-\frac{1}{2}|\hat{\sigma}_{z}|\frac{1}{2},-\frac{1}{2}\rangle
+(C^{\frac{1}{2}\frac{1}{2}}
_{10,\frac{1}{2}+\frac{1}{2}})^{2}
\langle\frac{1}{2},+\frac{1}{2}|\hat{\sigma}_{z}|\frac{1}{2},+\frac{1}{2}\rangle]\nonumber\\
&=&-\frac{M_{N}}{9m}
\end{eqnarray}
At last, the magnetic moment of the $uucd\bar{d}$ component should
be
\begin{eqnarray}
\mu&=&\frac{1}{6}(\frac{2M_{N}}{3m_{c}}+\frac{M_{N}}{m})-\frac{M_{N}}{9m}\nonumber\\
&=&(\frac{M_{N}}{18m}+\frac{M_{N}}{9m_{c}})
\end{eqnarray}

\section{the transition element between the $uuc$ and $uucd\bar{d}$ components}
\label{sec:c}

 As we know, on the hadron level, the baryon magnetic
form factor can be expressed as the matrix element of the vector
current operator in the following way (in Breit frame) \cite{wma}:
\begin{eqnarray}
\langle
P'|\vec{J}^{em}|P\rangle&=&\bar{u}(\vec{P}')\vec{\gamma}u(\vec{P})G_{M}(q^{2})\nonumber\\
\langle
P'|{J}_{0}^{em}|P\rangle&=&\bar{u}(\vec{P}')\gamma_{0}u(\vec{P})G_{M}(q^{2})
\end{eqnarray}
Here we get the relations between the electro-magnetic form factors
of the baryons and the matric elements of the current operators. In
the case of the magnetic form factor, we only need to consider the
first one. Assuming that both of the initial and final spin states
are $|1/2, \uparrow\rangle$, and
$\vec{q}=\vec{P}'-\vec{P}=q\vec{j}$, in the non-relativistic
approximation, the matrix element $\langle P'|\vec{J}^{em}|P\rangle$
will be
\begin{equation}
\langle
P'|\vec{J}^{em}|P\rangle=-\frac{i}{2M_{B}}q\hat{x}G_{M}(q^{2})
\label{basic}
\end{equation}
here $M_{B}$ denotes the mass of the baryon. Note that the vector
factor will be canceled in the final result. On the quark level,
taking the $qqqq\bar{q}$ components into account, the transition
matrix element $qqqq_{i}\bar{q}_{i}\rightarrow qqq\gamma$ will also
contribute to the baryon magnetic form factor. The transition
operator takes the following form:
\begin{eqnarray}
\hat{T}_{i}&=&Q_{i}\langle\vec{p}^{'}_{i}|\vec{j}_{i}
|\vec{p}_{i}\rangle_{(anni)}\,,\nonumber\\
&=&Q_{i}\hat{\sigma}_{i}\, .\label{ndia}
\end{eqnarray}
Analogous situation of the roper resonance decays to $N\gamma$ has
been considered in Ref. \cite{qbl}. As in equation (\ref{basic}), we
only need to calculate the x-component of the matrix elements of the
operator (\ref{ndia}). Note that the transition operator here
applies for all frames.

For the case of the $uucd\bar{d}\rightarrow uuc\gamma$ transition,
the calculation of the matrix element of the operator (\ref{ndia})
involves calculation of the overlap of the $uuc$ component with the
residual $uuc$ component that is left in the $uucd\bar{d}$ component
after the annihilation of a $d\bar{d}$ pair into a photon.

First, from equations (\ref{qqq})-(\ref{FS}), we can obtain that the
flavor-spin overlap factor may be
\begin{equation}
C_{FS}=\frac{\sqrt{2}}{4}
\end{equation}

The following step is the calculation of orbital matrix element.This
may be cast in the form
\begin{equation}
C_{O}=\langle\varphi_{00}(\vec{p}'_{1})\varphi_{00}(\vec{p}'_{2})
\varphi_{00}(\vec{p}'_{3})[111]_{C}|\delta(\vec{p}'_{1}-\vec{p}_{1})
\delta(\vec{p}'_{2}-\vec{p}_{2})\delta(\vec{p}'_{3}-\vec{p}_{3})
\delta(\vec{p}_{4}+\vec{p}_{5}-\vec{q})|\psi_{C}(\{\vec{p}_i\})\rangle,.
\label{co}
\end{equation}

The operator (\ref{ndia}) confines that the anti-quark should be in
the spin state $|1/2, \downarrow\rangle$, to lead to a nonzero
matrix element:
$\langle\bar{\chi}_{s_{z}}|\hat{\sigma}_{x}|1/2,\downarrow\rangle=-1$,
consequently, the CG coefficient $C^{\frac{1}{2}\frac{1}{2}}
_{1m,\frac{1}{2}s_{z}}$ in equation (\ref{wfc}) can only be the one
$C^{\frac{1}{2}\frac{1}{2}}
_{11,\frac{1}{2}-\frac{1}{2}}=\sqrt{\frac{2}{3}}$.

Note that only the orbital symmetry configuration of the
$uucd\bar{d}$ component that is described by $[31]_{X1}$ in equation
(\ref{CX}) leads to a nonzero matrix element, when multiplied with
the totally symmetric orbital state of the $uuc$ component upon
annihilation of the fourth quark with the fifth anti-quark in the
$uucd\bar{d}$ component. Consequently, we can obtain
\begin{eqnarray}
C_{O}&=&\langle\varphi_{00}(\vec{p}'_{1})\varphi_{00}(\vec{p}'_{2})
\varphi_{00}(\vec{p}'_{3})[111]_{C}|\delta(\vec{p}'_{1}-\vec{p}_{1})
\delta(\vec{p}'_{2}-\vec{p}_{2})\delta(\vec{p}'_{3}-\vec{p}_{3})
\delta(\vec{p}_{4}+\vec{p}_{5}-\vec{q})|\frac{1}{\sqrt{3}}\times\nonumber\\
&&\frac{3}{\sqrt{12}}
\varphi_{00}(\vec{p}_{1})\varphi_{00}(\vec{p}_{2})\varphi_{00}(\vec{p}_{3})\varphi_{1m}(\vec{p}_{4})
\varphi_{00}(\vec{p}_{5})\rangle\nonumber\\
&=&\frac{1}{4}C_{35}\frac{iq}{\omega_{5}}\exp[-\frac{q^{2}}{4\omega_{5}^{2}}]
\end{eqnarray}

Here we have obtained all the overlap factors, but we must notice
that the $d$ quark which annihilates with the anti-quark may also be
one of the first three quarks, for the $S_{4}$ symmetry, we only
need to multiply a factor $4$ to the overlap factors we have
obtained. On the other side, the process $uuc\gamma\rightarrow
uucd\bar{d}$ may contribute a same value as we have obtained, so
there is another factor $2$. Taking all the factors into account,
and eliminating the factor $-\frac{i}{2M_{\Sigma^{++}_{c}}}q\hat{x}$
which has been shown in equation (\ref{basic}), we can obtain
\begin{equation}
G^{anni}_{M}=-\frac{2\sqrt{3}}{9}C_{35}\sqrt{P_{3q}P_{(\Sigma_{c}^{++})d\bar{d}}}
\frac{M_{\Sigma^{++}_{c}}}{\omega_{5}}\exp[-\frac{q^{2}}{4\omega_{5}^{2}}]
\end{equation}
 The result is in units of the
$\frac{1}{2M_{\Sigma^{++}_{c}}}$. In the $q^{2}\longrightarrow 0$
limit, it will contribute to the baryon magnetic moment. In the
traditional way, we can express the result in units of the nuclear
magnetons as follow:
\begin{equation}
\mu_{\Sigma^{++}_{c}}=-\frac{2\sqrt{3}}{9}F_{35}
(P_{(\Sigma^{++}_{c})d\bar{d}})
\end{equation}

\end{appendix}

\end{document}